\documentclass[aps, prx, twocolumn, showpacs,superscriptaddress]{revtex4-1}
\usepackage{graphicx}
\usepackage{amsmath}
\usepackage{dsfont}
\usepackage{amssymb}
\usepackage{physics}
\usepackage{hyperref}
\usepackage{ulem}
\hypersetup{colorlinks=true,
	    final=true,
	    linkcolor=blue,
	    citecolor=blue,
	    filecolor=blue,
	    urlcolor=blue,
}

\newcommand{\LNO}[1]{La$_{3}$Ni$_{2}$O$_{#1}$}

\newcommand{\ef}{$\varepsilon_{\mathrm{F}}$}

\begin{document}

\title{Electronic structure and magnetic properties of \texorpdfstring{La$_{3}$Ni$_{2}$O$_{7}$}{La3Ni2O7} under pressure: active role of the Ni-\texorpdfstring{$d_{x^2-y^2}$}{dx2-y2} orbitals}
\author{Harrison LaBollita}
\email{hlabolli@asu.edu}
\affiliation{Department of Physics, Arizona State University, Tempe, AZ 85287, USA}
\author{Victor Pardo}
\affiliation{Departamento de Física Aplicada, Universidade de Santiago de Compostela, E-15782 Santiago de Compostela, Spain}
\affiliation{Instituto de Materiais iMATUS,
  Universidade de Santiago de Compostela, E-15782 Campus Sur s/n,
  Santiago de Compostela, Spain}
\author{Michael R. Norman}
\affiliation{  Materials Science Division, Argonne National Laboratory, Lemont, Illinois 60439, USA}
\author{Antia S. Botana}
\affiliation{Department of Physics, Arizona State University, Tempe, AZ 85287, USA}
\date{\today}

\begin{abstract}
Following the recent report of superconductivity in the bilayer nickelate La$_{3}$Ni$_{2}$O$_{7}$ under pressure, we present an analysis of the electronic and magnetic properties of \LNO{7} as a function of pressure using correlated density functional theory methods (DFT+$U$). At the bare DFT level, the electronic structure of the ambient and high-pressure phases of \LNO{7} are qualitatively similar. Upon including local correlation effects within DFT+$U$ and allowing for magnetic ordering, we find a delicate interplay between pressure and electronic correlations. Within the pressure-correlations phase space, we identify a region (at $U$ values consistent with constrained RPA) characterized by a high spin to low spin transition with increasing pressure. In contrast to previous theoretical work that only highlights the crucial role of the Ni-$d_{z^2}$ orbitals in this material, we find that the Ni-$d_{x^{2}-y^{2}}$ orbitals are active upon pressure and drive this rich magnetic landscape. This picture is preserved in the presence of oxygen deficiencies. 

\end{abstract}

\maketitle

The recent observation of superconductivity in low-valence layered nickelates has produced tremendous excitement in the community in the last few years, first in the infinite-layer compounds $R$NiO$_2$ ($R$ = rare-earth) \cite{Li2019superconductivity,Osada2020superconducting, Osada2021nickelate, Zeng2021superconductivity}, and more recently in the quintuple-layer compound Nd$_6$Ni$_5$O$_{12}$ \cite{Pan2021superconductivity}. Structurally, these materials possess quasi-two-dimensional NiO$_2$ planes (analogous to the  CuO$_2$ planes of the cuprates) and belong to a larger family represented by the general chemical formula $R_{n+1}$Ni$_{n}$O$_{2n+2}$ where $n$ denotes the number of NiO$_2$ planes per formula unit along the $c$ axis. The discovery of superconductivity in this family of nickel oxide compounds completed a long search to find materials that can serve as proxies for cuprate physics.

Despite many structural, chemical, and electronic similarities to the cuprates \cite{Botana2020similarities}, the superconducting layered nickelates show some relevant differences, the most obvious being their superconducting critical temperatures ($\sim$ 15 K) \cite{Osada2020phase, Osada2020superconducting, Osada2021nickelate, Zeng2021superconductivity}, which are much lower than those obtained in the cuprates. Improved crystalline quality samples of the infinite-layer nickelate \cite{Lee2022character}, as well as the application of hydrostatic pressure \cite{Wang2022Press} have shown small incremental increases in T$_{c}$, but still remain far away from typical cuprate values of T$_{c}\sim80-160$ K. 

Very recently, a breakthrough T$_{c}$ near 80 K  has been reported  in the bilayer Ruddlesden-Popper (RP) nickelate \LNO{7} at modest pressures (P $\sim 14-42$ GPa) \cite{sun2023superconductivity,hou2023emergence,zhang2023exps}. Bilayer \LNO{7} differs from the previously reported superconducting nickelates in that it belongs to the parent $R_{n+1}$Ni$_{n}$O$_{3n+1}$ series. As such, it possesses  NiO$_{6}$ layers (retaining the apical oxygens, rather than having a square planar environment for the Ni atoms). Also, the oxidation state of the Ni is 2.5+, corresponding to an average $3d^{7.5}$ filling, far from the hard-to-stabilize $\sim$$3d^{8.8}$ filling that all other previously observed superconducting layered nickelates present. 

Work on \LNO{7} prior to the discovery of superconductivity had focused on analyzing some of its structural \cite{zhang1994synthesis,Ling2000neutron} and electronic structure \cite{pardo2011metal} characteristics, suggesting that the emergence of a cuprate-like electronic structure could be possible in this compound. More recent theoretical studies aimed at exploring the electronic structure of \LNO{7} with pressure in relation to superconductivity (using a variety of techniques from DFT+$U$, to DFT+DMFT, GW+DMFT, and model Hamiltonians)  point instead to the active role of the Ni-$d_{z^2}$ states in the vicinity of the Fermi level, in contrast to the cuprates \cite{zhang2023electronic, chen2023critical, lechermann2023electronic, christiansson2023correlated, luo2023bilayer, gu2023effective, shen2023effective, zhang2023electronic, yang2023possible, liu2023swave, zhang2023structural, qu2023bilayer, yang2023minimal, zhang2023trends, lu2023superconductivity, tian2023correlation,vortex2023huang, jiang2023screening, liao2023correlations, liao2023interlayer,oh2023tj,qin2023singlets, sakakibara2023hubbard, Yang2023dmrg, shilenko2023, wu2023zhangrice}.

\begin{figure*}
    \centering
    \includegraphics[width=2\columnwidth]{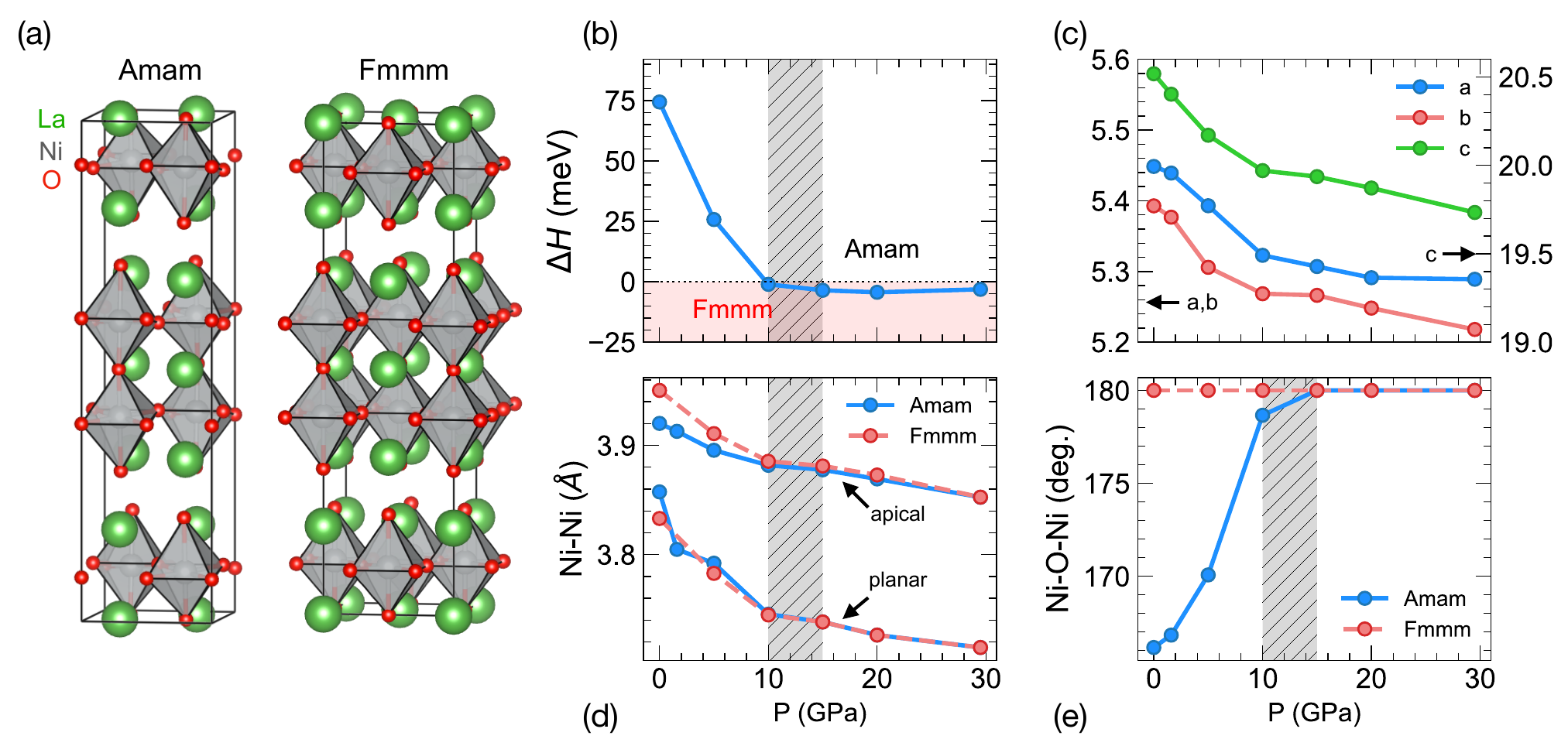}
    \caption{Crystal structures and structural data of \LNO{7} from DFT. (a) Crystal structure for \LNO{7} in the low-symmetry {\it Amam} (left) and high-symmetry {\it Fmmm} (right) phases. Small red spheres represent oxygen atoms, forming octahedral cages around the Ni (gray) atoms. The green spheres are the La atoms. Structural data for \LNO{7} as a function of pressure: (b) enthalpy ($H = E + PV$),  (c)  lattice constants extracted from the experimental data in \cite{sun2023superconductivity}, (d) relaxed apical and planar Ni-Ni bond lengths, and (e) relaxed Ni-O-Ni interplanar bond angles. The shaded, hatched area denotes the region where the structural transition occurs experimentally.}
    \label{fig:struct}
\end{figure*}

Here, we study the electronic structure of \LNO{7} under the influence of pressure using DFT and DFT+$U$ methods with different double-counting corrections. Within our methodology, we find a delicate interplay between pressure and electronic correlations with different phases closely competing in energy. 
Within the pressure-correlations phase diagram, we find that the region of $U$s consistent with constrained RPA (cRPA) calculations \cite{christiansson2023correlated} supports a spin-state transition with pressure. Specifically, we find a high-spin state at low pressures that transforms to a low-spin state at high pressures ($>$ 10 GPa).  Remarkably, we find that the ground states derived for all $U$s are controlled by the in-plane Ni-$d_{x^{2}-y^{2}}$ states that keep having an active role in this material despite the presence of Ni-$d_{z^2}$ states at the Fermi level. This active role of the $d_{x^{2}-y^{2}}$ states is preserved upon variations in the oxygen stoichiometry.

\begin{figure*}
    \centering  
    \includegraphics[width=2\columnwidth]{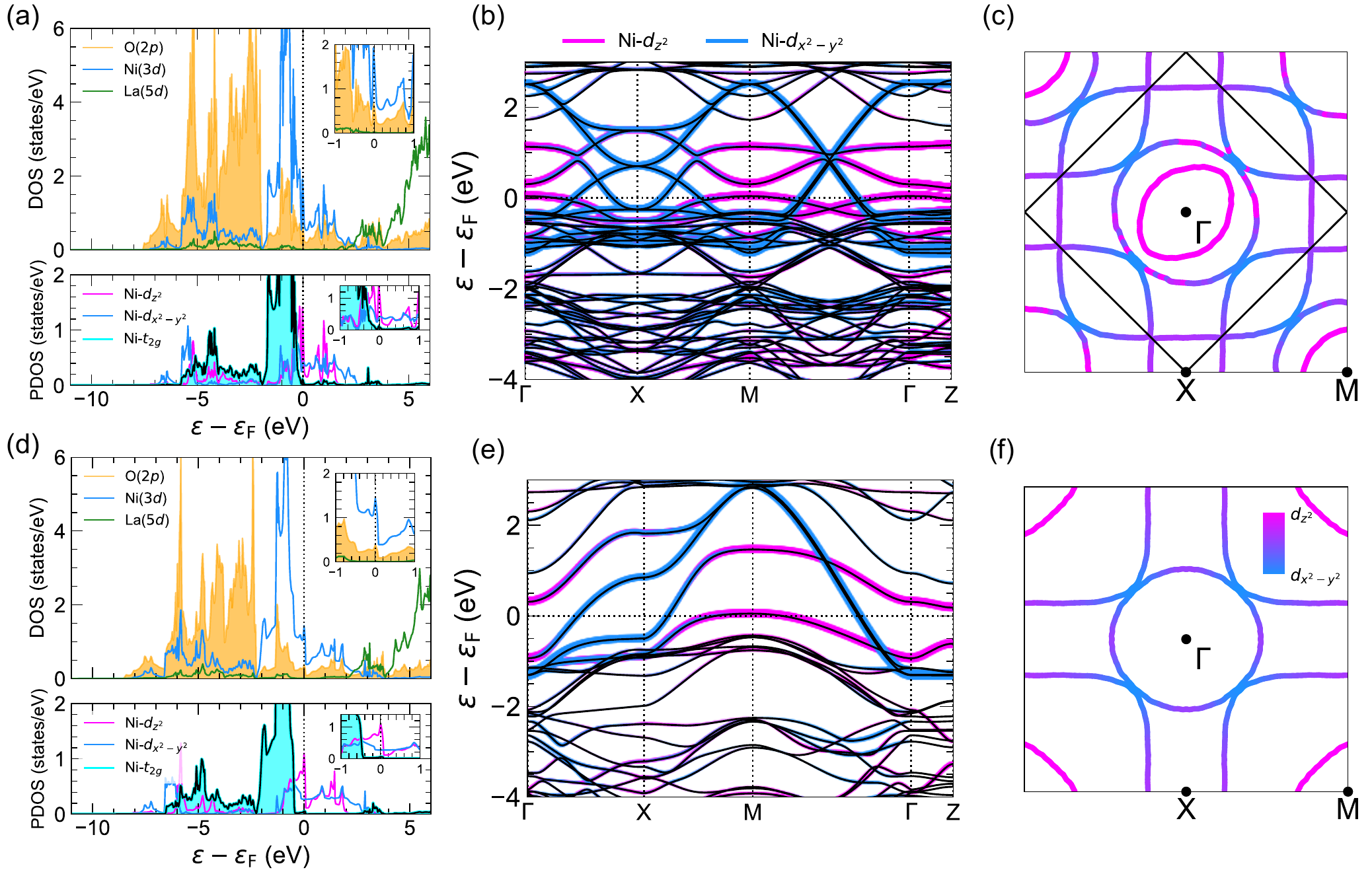}
    \caption{LDA non-magnetic electronic structure of \LNO{7} in the ambient pressure {\it Amam} phase (top row) and in the {\it Fmmm} phase at P = 29.5 GPa (bottom row). (a,d) Orbital resolved density of states (DOS) for the La, Ni, and O atoms (upper panels) and  partial density of states (PDOS) for the Ni-$d_{z^{2}}$, Ni-$d_{x^{2}-y^{2}}$, and Ni-$t_{2g}$ states (lower panels). Insets are a zoom-in  around the Fermi level. (b,e) Band structure along high-symmetry lines at ambient pressure ({\it Amam} structure) and at P= 29.5 GPa ({\it Fmmm} phase) where the corresponding $k$-path is the same in both zones (see more details in the Supplementary Note 3). Colors denote the orbital character for the Ni-$d_{z^{2}}$ and Ni-$d_{x^{2}-y^{2}}$ orbitals. (c,f)  Respective Fermi surfaces in the $k_{z}=0$ plane. Note the zone folding of \textit{Amam} relative to \textit{Fmmm}.}
    \label{fig:nm_es}
\end{figure*}

\section*{Results}

\subsection{Crystal structure}\label{sec:structure}

In addition to the report of superconductivity, the experiments performed in Ref. \onlinecite{sun2023superconductivity} reveal a structural transition in \LNO{7} from a low-pressure {\it Amam} phase to a high-pressure {\it Fmmm} phase. Hence, we start by carrying out structural optimizations under pressure for \LNO{7}  using first principles calculations. The experimental lattice parameters were adopted at the following pressures: P = 0, 5, 10, 15, 20, and 29.5 GPa \cite{Ling2000neutron, sun2023superconductivity}. With these lattice constants, the crystal structure for \LNO{7} was constructed in both the low-symmetry {\it Amam} and high-symmetry {\it Fmmm} space groups and the internal coordinates were fully relaxed.
In order to analyze the implications of the change in space group, we start by looking at the evolution of the {\it Amam} and {\it Fmmm} phases under pressure, as summarized in Figure \ref{fig:struct}. We find that, from DFT calculations, the {\it Amam} phase naturally evolves to the {\it Fmmm} phase when using the experimental lattice constants and relaxing the internal atomic coordinates. The {\it Fmmm} phase becomes energetically more favorable than the {\it Amam} phase at around $\sim$10 GPa, in agreement with experiments \cite{sun2023superconductivity}. Coinciding with this pressure, the octahedral NiO$_{6}$ tilting of the ambient pressure structure is (nearly) suppressed as shown by the Ni-O-Ni interplanar bond angle (see Fig.~\ref{fig:struct}), in agreement with other work \cite{geisler2023structure, zhang2023structural}.  Interestingly, if the crystal lattice is also allowed to relax, we find that contrary to initial X-ray diffraction (XRD) experiments, the material has a tendency to ``tetragonalize'' towards an I4/mmm structure with the $a$ and $b$ lattice constants collapsing to the same value (see Supplementary Note 1). Our results are in agreement with more recent XRD experiments performed at lower temperature that point towards the presence of an $I4/mmm$ structure in La$_3$Ni$_2$O$_7$ under pressure \cite{wang2023i4mmmexp}. %
Note that the suppression of the NiO$_{6}$ octahedra tilts still takes place in the relaxations allowing for both lattice constants and internal coordinates to be optimized. The discrepancy between the experimentally resolved and calculated structure could arguably also be due to the presence of oxygen deficiencies in the samples, consistent with transport data reporting a metal-to-insulator transition at low pressure \cite{sun2023superconductivity, zhang1994synthesis, filamentary}. We present a discussion on the potential role that variations in the oxygen stoichiometry may play in the electronic structure of \LNO{7} in the next section.

\subsection{\label{sec:results}Electronic structure and magnetism}

\subsubsection{\label{sec:para}Electronic structure at the LDA level}

A summary of the electronic structure of \LNO{7} at ambient and high (P = 29.5 GPa) pressure within LDA is presented in Fig.~\ref{fig:nm_es}. The electronic structure at ambient pressure (for the {\it Amam} structure) is characterized near the Fermi level by Ni-$e_{g}$ ($d_{z^{2}}$, $d_{x^{2}-y^{2}}$) states hybridized with O($2p$) states, which is consistent with previous works \cite{zhang2023electronic,lechermann2023electronic,sun2023superconductivity,yang2023orbitaldependent}. The Ni-$t_{2g}$ states are completely occupied and centered around $-2$ eV, just above the top of the O($2p$) bands. The rare-earth La($5d$) states are completely removed from the low-energy physics of this material and are unoccupied, unlike the superconducting infinite-layer and quintuple-layer nickelates, where the role of the rare-earth band degree of freedom is still being highly contested \cite{Lechermann2020multiorbital, Botana2020similarities, Karp2020comparative, Karp2020manybody, Goodge2020doping, Petocchi2020normal, Louie2022twogap, Labollita2022manybody, Labollita2023conductivity}. The removal of the La($5d$) states from the vicinity of the Fermi level is expected given the change in nominal valence for the Ni ions from 1.2+ to 2.5+.
The derived charge-transfer energy ($\Delta = \varepsilon_{d} - \varepsilon_{p}$) is  $\Delta = 3.2$ eV, much reduced from that in layered nickelates \cite{Nica2020theoretical, Botana2020similarities, Labollita2021electronic} and closer to a typical cuprate value. 

Focusing on the Ni-$e_{g}$ states, the Ni-$d_{z^2}$ states are split by $\sim 1$ eV into an occupied bonding and unoccupied antibonding combination due to the quantum confinement of nickel oxygen bilayers in the structure \cite{pardo2012pressure, Pardo2010quantum,PardoPRB2011}. The presence of apical oxygens broadens the Ni-$d_{z^2}$ bands with respect to the low-valence layered compounds such as \LNO{6} or La$_4$Ni$_3$O$_8$, but the bonding-antibonding splitting caused by the blocking layers is still present in \LNO{7} \cite{jung2022RPs}. 
The Ni-$d_{x^{2}-y^{2}}$ dispersion is large with a bandwidth of $\sim$2.5 eV and this orbital remains only partially occupied. As mentioned above, nominally, the Ni valence would be Ni$^{2.5+}$ ($3d^{7.5}$). As the $t_{2g}$ electronic states are completely occupied, this average filling means that 1.5 $e_g$-electrons per Ni need to be accommodated close to the Fermi level given that the O($2p$) bands are (almost) completely filled in this material.

Turning to the electronic spectrum of \LNO{7} at high pressure (P = 29.5 GPa, $Fmmm$ phase), we find that the overall electronic structure within LDA is qualitatively similar to the ambient pressure {\it Amam} phase, with some quantitative differences. The Ni $d_{x^{2}-y^{2}}$ dispersion increases to 4 eV, the bonding-antibonding Ni-$d_{z^{2}}$ splitting increases to $\sim$ 1.5 eV, and the charge-transfer energy $\Delta$ value increases to $3.6$ eV. Also, the Ni-$t_{2g}$ bands are pushed farther away from the Fermi level. Similar to the ambient pressure case, the dominant DOS around the Fermi level (\ef{}) for the non-magnetic calculation at 29.5 GPa is that coming first from Ni-$d_{z^2}$ orbitals followed by that of the Ni-$d_{x^{2}-y^{2}}$ orbitals.

The corresponding LDA Fermi surfaces at ambient pressure and at P= 29.5 GPa are shown in Figs.~\ref{fig:nm_es}(c,f), respectively. Both surfaces are comprised of two sheets from hybridized Ni-$e_{g}$ bands (an electron sheet centered at $\Gamma$ and a larger hole sheet centered around $M$) and small hole pockets at $M$ coming from the flattened Ni-$d_{z^{2}}$ bands. The zone folding between the {\it Amam} and {\it Fmmm} Brillouin zones is clearly seen from the Fermi surfaces (see the Supplementary Note 3 for more details on the corresponding Brillouin zones). We explore possible instabilities of the Fermi surface at the LDA level by calculating the static susceptibility from the near Fermi level bands (see Supplementary Note 4). Interestingly, we find there are no obvious
trends for nesting. This is in contrast to previous tight-binding calculations of the static susceptibility~\cite{zhang2023structural}.

\begin{figure}
    \centering
    \includegraphics[width=\columnwidth]{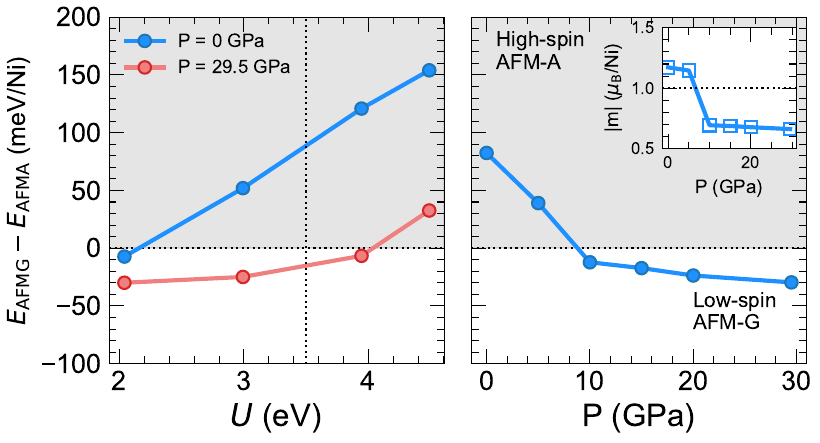}
    \caption{Energetics of different magnetic states in \LNO{7}. Left panel: Energy difference between high-spin A-type AFM (AFM-A) and low-spin G-type AFM (AFM-G) phases at P = 0 ({\it Amam} phase) and 29.5 GPa ({\it Fmmm} phase) as a function of $U$. Right panel: Energy difference between AFM-A and AFM-G phases as a function of pressure for $U = 3.5$ eV. Hund's coupling ($J_{\mathrm{H}}$) is fixed to 0.7 eV. The inset shows the evolution of the Ni magnetic moment ($\mu_{\mathrm{B}}$) as a function of pressure.}
    \label{fig:interplay}
\end{figure}

\subsubsection{\label{sec:groundstates}Interplay between pressure and electronic correlations}
Experiments suggest a complicated electronic and magnetic structure for \LNO{7}. Transport and magnetic susceptibility measurements hint at charge and spin ordering similar to the trilayer La$_{4}$Ni$_{3}$O$_{10}$ Ruddlesden-Popper nickelate, and point to the presence of antiferromagnetic correlations \cite{Liu2022Evide}. Recent muon spin relaxation ($\mu$SR) data show evidence for long-range magnetic order,  consistent with spin density wave formation \cite{chen2023musr}.

On this note, we explore the pressure and correlation phase space to reveal magnetic tendencies in \LNO{7} within an LDA+$U$ framework. We note that a careful choice
of the DFT+$U$ implementation is important to capture the physics of
these nickelates \cite{Zhang2017large}; for instance, when the two spin-states are nearly degenerate, it becomes essential when applying DFT+$U$ to understand the tendencies of the various choices of double counting correction: the around mean field (AMF) scheme is known to favor the
stabilization of low-spin configurations, whereas the fully localized
limit (FLL) favors high-spin configurations \cite{Pickett2009anisotropy}. The choice of AMF double counting has been shown in the past to give a reliable comparison to experiments in other layered nickelates \cite{Zhang2017large,Krishna2020effects,Botana2016charge, Botana2017electron,pardo2012pressure} with FLL giving rise to ground states that are inconsistent with experiments \cite{Zhang2017large}. Hence, we focus on the AMF scheme in the main text and present LDA+$U$(FLL) results in Supplementary Note 5A. For each double-counting scheme, we study three possible simple magnetic orderings within a range of $U$'s from 1 to 5 eV: ferromagnetic (FM), A-type antiferromagnetic (AFM-A), and G-type antiferromagnetic (AFM-G), where AFM-A(G) corresponds to Ni moments coupled FM (AFM) in-plane and AFM (AFM) out-of-plane (see Supplementary Note 5A for more details). 
We note that the inclusion of a Hubbard $U$ is necessary to initially converge the magnetically ordered states, otherwise, a non-magnetic solution is obtained. 

\begin{figure*}
    \centering
    \includegraphics[width=2\columnwidth]{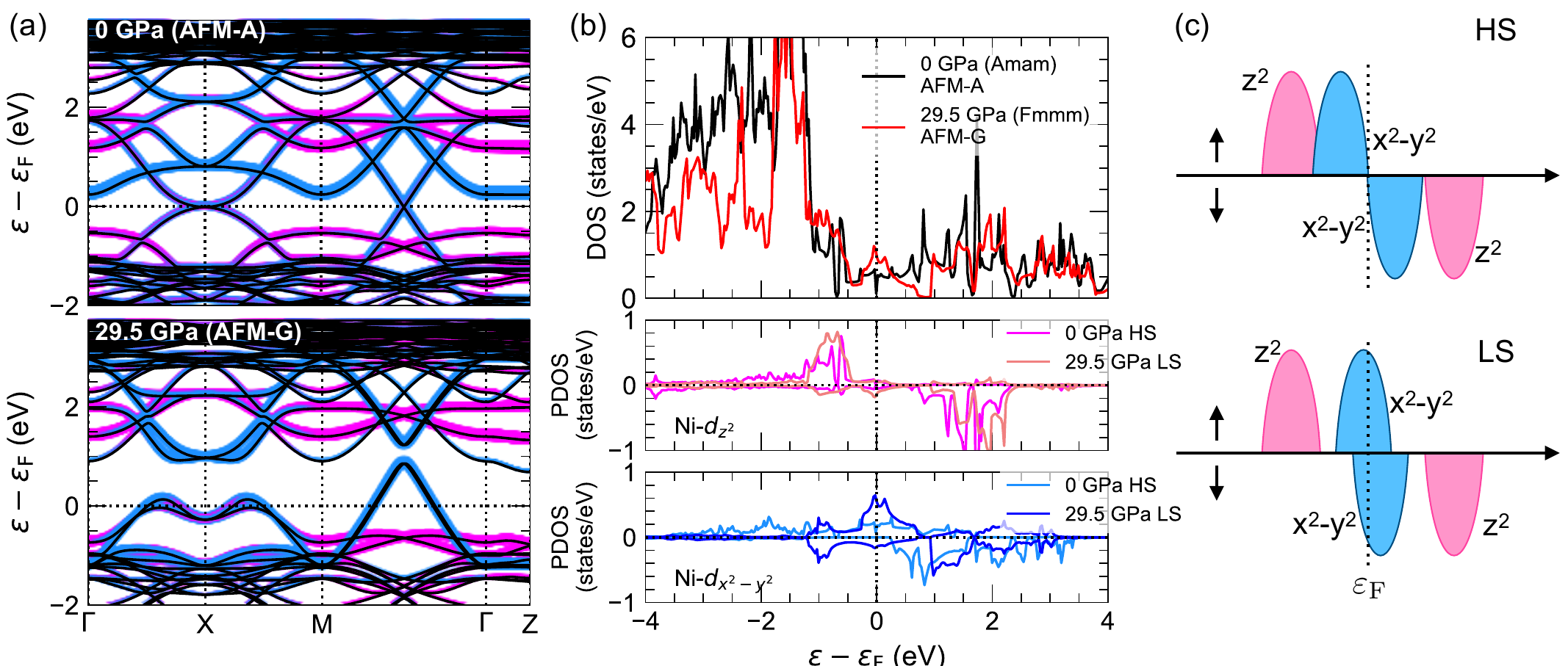}
    \caption{LDA+$U$(AMF) ($U=3.5$ eV, $J_{\mathrm{H}}=0.7$ eV) spin-polarized electronic structure of \LNO{7} under pressure. (a) Orbital-resolved (Ni-$e_{g}$) band structures along high-symmetry lines in the Brillouin zone for \LNO{7}: P = 0 GPa {\it Amam} structure in its HS AFM-A ground state (top) and P = 29.5 GPa {\it Fmmm} structure in its LS AFM-G ground state (bottom), where blue (pink) corresponds to Ni-$d_{x^{2}-y^{2}}$ (Ni-$d_{z^{2}}$) character. Note for the \textit{Fmmm} structures (29.5 GPa), supercells need to be constructed to allow for G-type ordering, which have \textit{Amam} symmetry. (b) Total density of states (DOS) (top) and spin-and orbital-resolved Ni-$e_{g}$ partial density of states (PDOS) (bottom) for the ambient (AFM-A) and high pressure (P = 29.5 GPa; AFM-G) phases. (c) Schematic Ni-$e_{g}$ diagram for the high-spin (top) and low-spin (bottom) solutions.}
    \label{fig:mag-es}
\end{figure*}

Figure \ref{fig:interplay} summarizes two representative cuts from the pressure and correlation phase space within LDA+$U$(AMF). From the energy difference between the different magnetic configurations as a function of Hubbard $U$, we find a transition from a high-spin (HS) AFM-A state at ambient pressure to a low-spin (LS) AFM-G  phase with pressure for a range of $U$ values between 2 and 4 eV.  
Above $\sim4$ eV, this transition is suppressed and the ground state remains in the HS AFM-A phase at all pressures. This competition of different magnetically ordered phases with different magnetic moments and exchange couplings suggests there is a rich energy landscape with many (nearly) degenerate ground states. 

With the two distinct $U$-dependent regimes identified, we now explain how the electronic structure of \LNO{7} evolves as a function of pressure. In the high-$U$ regime, the HS AFM ground state at all pressures portrays a van Hove singularity pinned at the Fermi level which would make the AFM order unstable (see Supplementary Note 5B). For this reason, we focus here on the region of the pressure-correlations phase space in which a spin-state transition takes place (for 2 eV $< U < $4 eV), that is consistent with the $U$= 3.4 eV derived for \LNO{7} from cRPA~\cite{christiansson2023correlated}. As shown in Fig.~\ref{fig:interplay}, at this value of $U$ the spin-state transition from a HS to a LS state Ni occurs at around 10 GPa, which corresponds to the structural transition from the {\it Amam} to the {\it Fmmm} phase. In the LDA+$U$(FLL) scheme, this spin-state transition still occurs but within a narrower range of $U$ values (see more details in Supplementary Note 5A).

The evolution of the band structure across the spin-state transition is shown in Fig.~\ref{fig:mag-es}(a) where colors denote the weights of the Ni-$d_{z^{2}}$ and Ni-$d_{x^{2}-y^{2}}$ orbitals. At ambient pressure in the HS ground state, we find that the near Fermi level bands correspond to hybridized Ni-$e_{g}$ ($d_{z^{2}}$ and $d_{x^{2}-y^{2}}$) states with a van Hove singularity at the X point essentially at the Fermi energy (the doubled bands are due to $Amam$ zone folding). The bonding Ni-$d_{z^{2}}$ states are fully occupied. The Ni-$d_{x^{2}-y^{2}}$ majority states are nearly full, with a $\sim 0.75$ filling obtained from a simple integration of the occupied spectrum. The Ni magnetic moment is $\sim$ 1.2$\mu_{\mathrm{B}}$, which is slightly reduced from the HS ionic value due to hybridization effects. The overall weak metallic signature in this phase coincides with the poor conductivity evidenced by the experimental resistivity \cite{sun2023superconductivity} at ambient pressure.  

Near the experimentally reported structural transition (P = 10 GPa), we find that the electronic structure changes dramatically with the material adopting a LS ground state for the Ni atoms characterized by the dominant role of the $d_{x^{2}-y^{2}}$ states around the Fermi level. We find that with pressure the Ni-$d_{x^{2}-y^{2}}$ orbital becomes half-filled per spin channel leading to an overall quarter-filling, which further stabilizes the in-plane AFM coupling~\cite{pardo2011metal}. As shown in Fig.~\ref{fig:struct}, applying pressure decreases the planar Ni-Ni distance faster than the apical Ni-Ni distance. Decreasing the planar Ni-Ni distance favors the LS state to accommodate for this reduction \cite{Matsumoto2019}. We note a similar spin-state transition has been reported in the trilayer nickelate La$_{4}$Ni$_{3}$O$_{8}$ under pressure \cite{pardo2012pressure}. 

In Fig.~\ref{fig:mag-es}(b) we compare the total density of states (DOS) and Ni-$e_{g}$ partial density of states (PDOS) for the high-spin and low-spin solutions at the same the two pressure extremes: 0 GPa (HS) and 29.5 GPa (LS). Overall, it can be clearly seen that the PDOS is dominated by Ni-$d_{x^2-y^2}$ states around the Fermi level for all solutions. Analyzing the Ni-$d_{z^{2}}$ PDOS, their spectrum remains inert between the LS and HS solutions at all pressures.  For the Ni-$d_{x^{2}-y^{2}}$ orbitals, we find that a large redistribution takes place when comparing the HS solution (ground state at 0 GPa) and the LS solution (ground state above 10 GPa), indicating that these are the orbitals that drive the spin-state transition. This is also revealed in Supplementary Note 5C  from the (nearly) constant Ni-$d_{z^{2}}$ occupations under varying pressure across different spin-state solutions, in contrast to the notable variations in Ni-$d_{x^{2}-y^{2}}$ occupations between spin-state solutions. Further, our analysis of the changes in the electronic structure of \LNO{7} induced by oxygen deficiencies also reveals a dominant role of the $d_{x^2-y^2}$ states
around the Fermi level (see Supplementary Note 2 for further details).  
 
 Based on this derived spin-state transition with very clear changes in the electronic structure with pressure, we speculate that the HS (ambient pressure) solution is likely unfavorable for superconductivity in \LNO{7}. Importantly, the LS ground state obtained at high pressure (where superconductivity arises) that is dominated by Ni-$d_{x^{2}-y^{2}}$ orbitals suggests that the physics \LNO{7} could share similarities with the high-$T_{c}$ cuprates.

\section*{\label{sec:summary}Summary and discussion}
We have studied the evolution of the electronic structure of the bilayer RP nickelate \LNO{7} with pressure using a correlated density-functional theory framework. We capture the experimentally observed structural transition from \textit{Amam} to \textit{Fmmm} (which corresponds to the suppression in the tilts of the NiO$_{6}$ octahedra) and find a possible transition to a tetragonal \textit{I4/mmm} space group that preserves the suppression of octahedral tilts. At the LDA level, the electronic structure of \LNO{7} at ambient and high-pressures is qualitatively similar: hybridized Ni-$e_{g}$ ($d_{z^2}$+$d_{x^{2}-y^{2}}$) states are dominant near the Fermi level with additional weight from the O($2p$) orbitals without the involvement of rare-earth bands in the low-energy physics, in contrast to the superconducting (infinite-)layered nickelates. 

Using LDA+$U$, we explore the pressure and correlation phase space and find crucial differences with respect to the uncorrelated electronic structure. At low $U$ values (between 2 and 4 eV, consistent with cRPA) a spin-state (HS to LS) transition with pressure takes place. This transition is driven by a redistribution of the (dominant) Ni-$d_{x^{2}-y^{2}}$ orbitals. The active role of the Ni-$d_{x^{2}-y^{2}}$ orbitals is preserved upon variations in oxygen stoichiometry. Based on our derived spin-state transition in \LNO{7}, we hypothesize that the HS solution is unfavorable for superconductivity in this material while the LS solution may promote it instead. Overall, we conclude that in the bilayer RP nickelate the $d_{x^2-y^2}$ states dominate the low energy physics, with the $d_{z^2}$ states acting largely as spectator-like in nature.

\section*{\label{sec:methods}Methods}
For the structural relaxations the internal coordinates of the atomic positions were optimized using the plane-wave pseudopotential DFT code VASP \cite{Kresse:1993bz, Kresse:1999dk, Kresse:1996kl} within the generalized gradient approximation (GGA) \cite{gga_pbe}. The number of plane waves in the basis was set by an energy cutoff of 500 eV. The integration in reciprocal space was carried out on an $8\times8\times4$ grid. The internal forces on each atom were converged to $10^{-6}$ eV/\AA{}.

The electronic structure for each structure was then calculated within DFT as implemented in the all-electron, full-potential code {\sc wien2k} \cite{Blaha2020wien2k}. The local-density approximation (LDA) \cite{perdew1992accurate} was adopted for the exchange-correlation functional and correlation effects were included using DFT + Hubbard $U$ (DFT+$U$), which allows for the incorporation of local Coulomb interactions for the localized Ni($3d$) states. Within the DFT+$U$ scheme, the double-counting term plays a central role in determining the underlying low-energy physics of the material~\cite{Pickett2009anisotropy}. We have adopted two choices that are commonly used for the double-counting correction: the fully-localized limit (FLL)~\cite{ldau_fll} and the around mean-field (AMF) \cite{ldau_amf}.  The results shown in the main text adopt the AMF double counting term. We use a range of $U$ values from 1 to 5 eV. The Hund's coupling $J_{\mathrm{H}}$ is fixed to a typical value of 0.7 eV for transition-metal $3d$ electrons. A 10$\times$10$\times$9 and a 10$\times$10$\times$10 $k$-point mesh was used for Brillouin zone integration for the \textit{Amam} phase and for the \textit{Fmmm} phase, respectively. The basis set size is determined by $RK_{\mathrm{max}} = 7$ and muffin-tin radii (in atomic units) set to 2.30, 1.86, and 1.65 for La, Ni, and O, respectively.

\section*{author contributions}
ASB conceived the study. HL performed the first-principles calculations. MRN performed the susceptibility calculations. All authors contributed equally to discussing and writing the manuscript.

\section*{acknowledgments}
 HL and ASB acknowledge the support from NSF-DMR 2045826 and from the ASU Research Computing Center for HPC resources. VP acknowledges support from the Ministry of Science of Spain through the Project No. PID2021-122609NB-C22. MRN was supported by the Materials Sciences and Engineering Division, Basic Energy Sciences, Office of Science, U.S.~Dept.~of Energy.

\bibliography{ref.bib}

\end{document}


\title{Supplemental Information for ``Electronic structure and magnetic properties of \texorpdfstring{La$_{3}$Ni$_{2}$O$_{7}$}{La3Ni2O7} under pressure: active role of the Ni-\texorpdfstring{$d_{x^2-y^2}$}{dx2-y2} orbitals''}
\author{Harrison LaBollita}
\email{hlabolli@asu.edu}
\affiliation{Department of Physics, Arizona State University, Tempe, AZ 85287, USA}
\author{Victor Pardo}
\affiliation{Departamento de Física Aplicada, Universidade de Santiago de Compostela, E-15782 Santiago de Compostela, Spain}
\affiliation{Instituto de Materiais iMATUS,
  Universidade de Santiago de Compostela, E-15782 Campus Sur s/n,
  Santiago de Compostela, Spain}
\author{Michael R. Norman}
\affiliation{  Materials Science Division, Argonne National Laboratory, Lemont, Illinois 60439, USA}
\author{Antia S. Botana}
\affiliation{Department of Physics, Arizona State University, Tempe, AZ 85287, USA}

\maketitle

 \section*{\label{sec:addstruct}Supplementary Note 1: ``Tetragonalization'' of crystal structure}
In the main text we build the crystal structure of \LNO{7} using the experimental lattice parameters at different pressures. Additionally, we investigated {\it ab-initio} predictions of the crystal structure when pressure is applied. This is achieved by relaxing the crystal lattice and internal coordinates where an additional term is added to the total energy calculation to incorporate the stress tensor modeling the external pressure. Structural optimizations are completed in the VASP code using the same computational settings described in the Methods section for the structural relaxations.

\begin{figure}
     \centering
     \includegraphics[width=0.35\columnwidth]{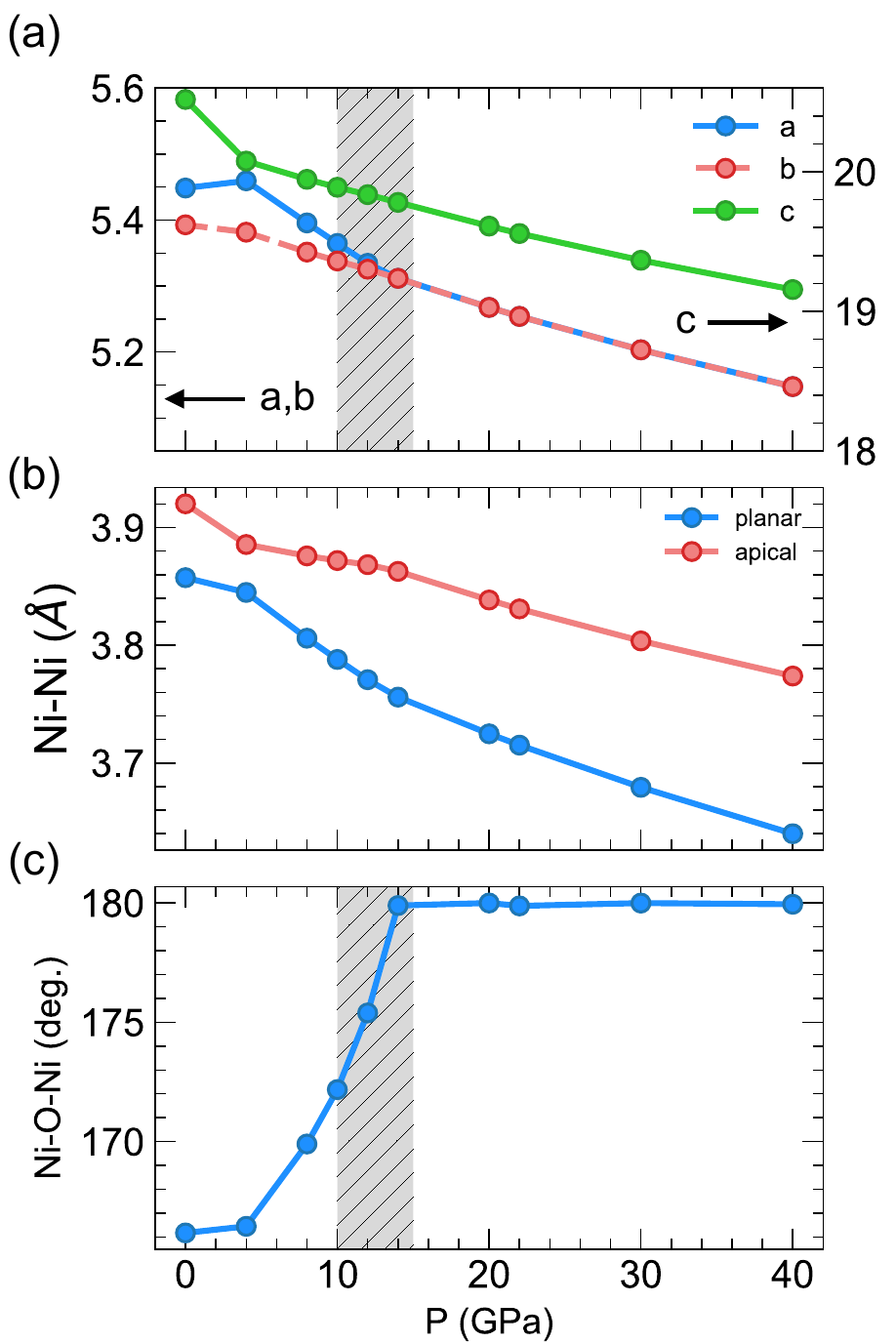}
     \caption{Structural data from {\it ab-inito} relaxations that allow for the lattice and internal coordinates to be optimized: (a) lattice constants, (b) Ni-Ni distances (apical and planar), and (c) Ni-O-Ni interplanar bond angle. Shaded regions denote the experimental region where a structural transition occurs.} 
     \label{fig:cmp_tetra}
 \end{figure}

The structural data from our fully optimized crystal structures are summarized in Fig.~\ref{fig:cmp_tetra}. In contrast to the experimental data, we find that the in-plane lattice constants ($a$, $b$) ``tetragonalize'' collapsing to the same value. We hypothesize that a possible explanation for the mismatch between our results and experiment could be the presence of oxygen deficiencies in the samples, known to be a challenge in the bilayer Ruddlesden-Popper nickelate \cite{zhang1994synthesis}.
Importantly, even in this (quasi)-tetragonal structure, we find that the main signatures of the experimental structure under pressure remain with the Ni-Ni distances and Ni-O-Ni interplanar angle following the same trends (see Fig.~\ref{fig:cmp_tetra}(b,c)). 

\begin{figure}
    \centering
    \includegraphics[width=0.45\columnwidth]{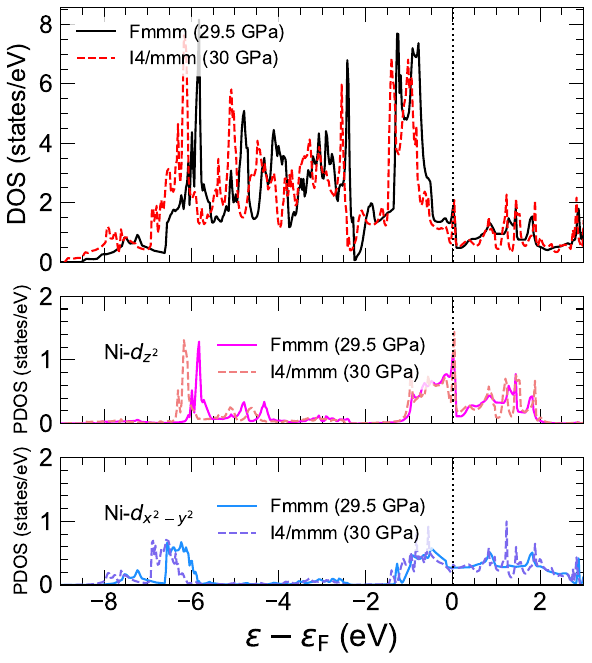}
    \caption{Comparison of the non-magnetic LDA electronic structure of \LNO{7} at 29.5 GPa in the {\it Fmmm} phase using the experimental lattice parameters and the fully DFT-relaxed \LNO{7} structure at 30 GPa ({\it I4/mmm}). Total DOS (top), Ni-$d_{z^{2}}$ PDOS (center), and Ni-$d_{x^{2}-y^{2}}$ PDOS (bottom).} 
    \label{fig:cmp_tetra_dos}
\end{figure}

Figure~\ref{fig:cmp_tetra_dos} compares the LDA density of states (DOS) of \LNO{7} obtained using the experimental lattice parameters at P = 29.5 GPa ({\it Fmmm} phase) and our fully relaxed {\it ab-initio}-determined crystal structure at 30 GPa. The electronic structure for both settings is essentially identical and only small quantitative differences can be noticed.

\section*{\label{sec:Ov}Supplementary Note 2: Doping via oxygen vacancies}
The experiments in Ref. \cite{sun2023superconductivity} show a metal-to-insulator transition in \LNO{7} when going from ambient pressure to 1 GPa. Previous work in the La$_3$Ni$_2$O$_{7-\delta}$ series \cite{zhang1994synthesis} showed a very similar trend when comparing stoichiometric \LNO{7} with oxygen-deficient samples.  Thus, we analyze how the electronic structure is affected by electron doping induced by oxygen non-stoichiometries, in particular whether the dominance of $d_{x^2-y^2}$ states around the Fermi level described in the main text remains in the oxygen-deficient compound.

\begin{figure}
    \centering
    \includegraphics[width=0.45\columnwidth]{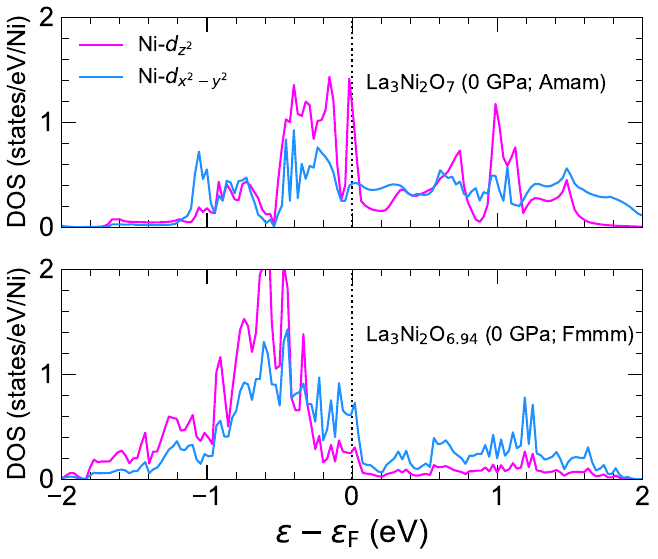}
    \caption{Ni-$e_{g}$ density of states (DOS) for stoichiometric \LNO{7} at ambient pressure ({\it Amam}) (top) and 
    oxygen-deficient \LNO{6.94}, which was resolved in the {\it Fmmm} space group at ambient pressure \cite{park2001xanes} (bottom). For \LNO{6.9375} the Ni-$e_{g}$ DOS has been averaged over all inequivalent Ni sites.} 
    \label{fig:Ovdope}
\end{figure}

Experimentally, the structure of \LNO{6.94} has been previously resolved in the {\it Fmmm} space group \cite{park2001xanes}. Using the experimental structural data, we have constructed a 4 $\times$ 4 $\times$ 1 supercell 
in which a single O atom can be removed to obtain the desired stoichiometry (\LNO{6.9375}) corresponding to around 2\% electron doping. As shown in Ref. \onlinecite{PardoPRB2011}, the removal of an apical oxygen is most probable based on the calculated energetics. We allow the internal atomic coordinates to relax to enable the lattice to respond to this local defect using the same procedures detailed in the Methods section of the main text. Because this calculation is computationally intensive, we analyze only the non-magnetic state at the level of LDA. The averaged Ni-$e_{g}$ partial density of states (PDOS) for \LNO{7} and \LNO{6.94} at ambient pressure are shown in Fig.~\ref{fig:Ovdope}. As described above, within LDA, the states populating the Fermi level at any pressure in \LNO{7} are the strongly mixed Ni-$e_{g}$ states with the Ni-$d_{z^{2}}$ showing a large peak. Interestingly, for the oxygen-deficient sample, we find that the states derived from the Ni $d_{x^{2}-y^{2}}$ orbitals become largely dominant at the Fermi level already in the LDA results at ambient pressure. This type of calculation suggests that oxygen vacancies should play an important role in the electronic structure of \LNO{7} and will require further theoretical and experimental investigation.

\section*{\label{sec:zones}Supplementary Note 3: Brillouin zone coordinates}
Throughout this work, we compare the electronic structure of \LNO{7} within two different space groups {\it Amam} and {\it Fmmm}. For all calculations the high-symmetry points are given by: $\Gamma = (000)$, $X = (110)$, $M = (020)$, and $Z = (002)$ in units of ($\pi$/a, $\pi/b$, $\pi/c$) and assuming the long axis is along $c$. Table~\ref{tab:spacegroup} summarizes the different space groups of the structures used in this work.  

\begin{table}[h]
    \centering
    \begin{tabular*}{\columnwidth}{l@{\extracolsep{\fill}}ccc}
    \hline\hline
    Calculation &  Pressure (GPa) & Space group & Figure\\
    \hline
    NM          &      0    &  {\it Amam} & 2\\
    AFM-A       &      0    &  {\it Amam} &  S6\\
    AFM-A  (HS) &      10   &  {\it Fmmm} & S6\\
    NM          &      29.5   &  {\it Fmmm} & 2\\
    AFM-A  (HS) &      29.5   &  {\it Fmmm} & S6\\
    AFM-G  (LS) &      29.5   &  {\it Amam} & 4\\
    \hline\hline
    \end{tabular*}
    \caption{Summary of space groups used in each of the calculations presented in the manuscript. Note, for the 10 GPa and 29.5 GPa AFM-G calculations, we construct a supercell from the {\it Fmmm} structures to allow for G-type ordering, and this cell has {\it Amam} symmetry.}
    \label{tab:spacegroup}
\end{table}

\section*{\label{sec:instability}Supplementary Note 4: Fermi surface and instabilities}
\begin{figure*}
    \centering
    \includegraphics[width=0.9\columnwidth]{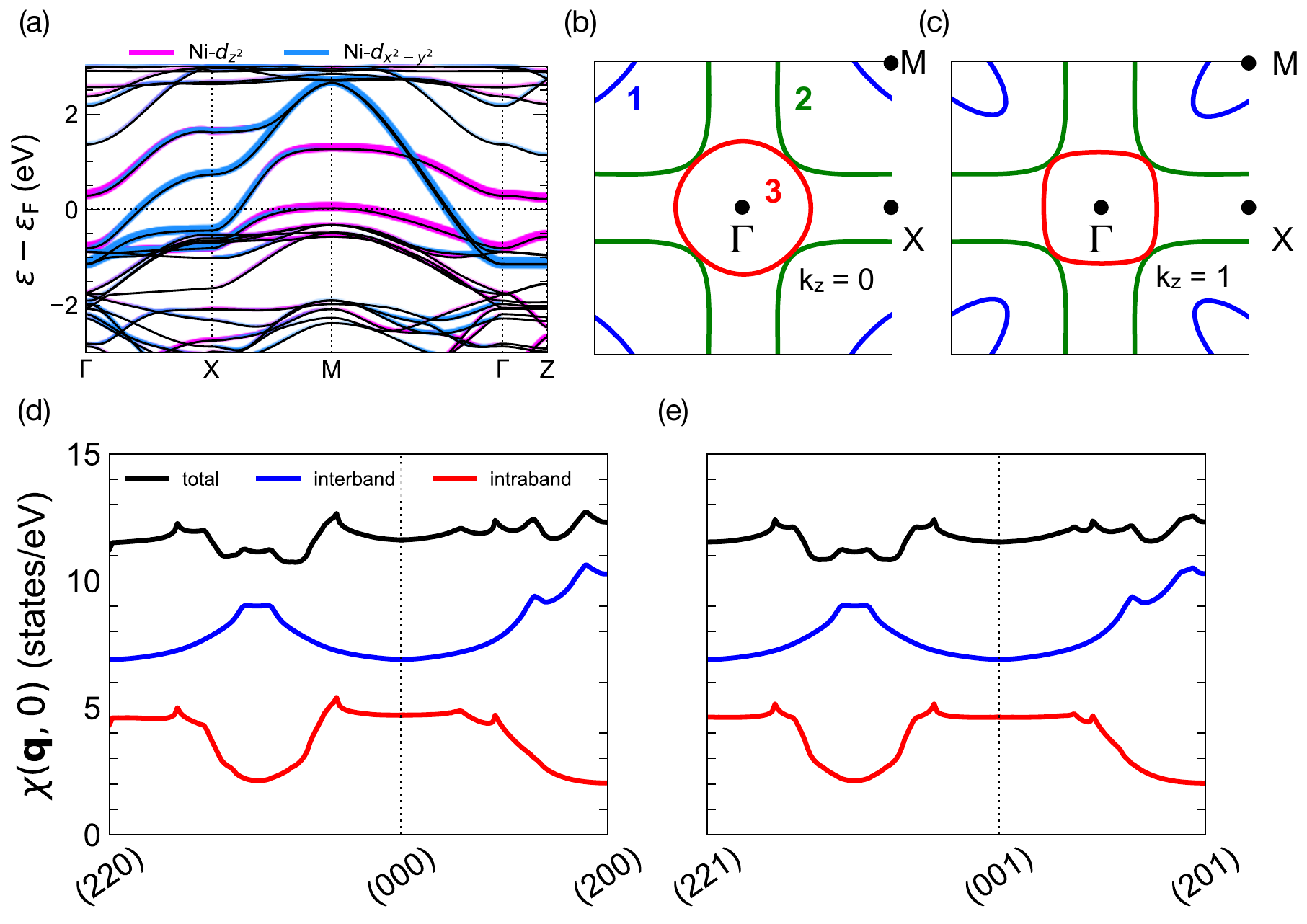}
    \caption{DFT static susceptibility $\chi_{nn'}({\bf q},0)$ for \LNO{7}. (a) Band structure along high-symmetry lines within the fat band representation (Ni-$e_{g}$ orbitals highlighted) for \LNO{7} using {\it Fmmm} coordinates at ambient pressure. LDA Fermi surfaces in the (b) $k_{z}=0$ and the (c) $k_{z}=\pi$/c planes. Colors correspond to band indices 1,2,3 used in the susceptibility calculation. Static susceptibility $\chi_{nn'}({\bf q},0)$ along high-symmetry directions for (d) $q_{z}=0$ and (e) $q_{z}=\pi$/c. High-symmetry coordinates are shown in Supplementary Section III and the coordinates for the susceptibility plots are in $\pi$/a,b,c units.}
    \label{fig:suscept}
\end{figure*}

To gain further insight into possible Fermi surface instabilities, we calculate the static susceptibility using the LDA non-magnetic band structure. We start by performing a paramagnetic calculation within \LNO{7} using {\it Fmmm} coordinates at ambient pressure. The obtained LDA bands are shown in Fig.~\ref{fig:suscept}(a) where the Ni-$e_{g}$ orbital characters are highlighted in color. The corresponding LDA Fermi surface in the $k_{z}=0$ and $k_{z}=\pi/c$ planes are shown in Fig.~\ref{fig:suscept}(b,c). The Fermi surface is comprised of three sheets: (1) small hole pockets at the zone corners from the bonding Ni-$d_{z^{2}}$ band, (2) large hole pockets centered at the zone corners that extend almost to $X$, and (3) an electron pocket centered around $\Gamma$ with mostly Ni-$d_{x^{2}-y^{2}}$ character. Comparing the two $k_{z}$ cuts, we can see there is noticeable $k_{z}$ dispersion. 

To calculate the static susceptibility $\chi_{nn'}({\bf q},\omega=0)$ from the LDA bands (subscripts are the band indices), we interpolate the near Fermi level bands using a Fourier spline series. Specifically, a Fourier series spline fit \cite{KOELLING1986253} to the DFT bands was made with 2813 face centered orthorhombic ({\it Fmmm}) Fourier functions fit to 511 $k$-points in the irreducible wedge of the Brillouin zone. Both the density of states and the susceptibility were calculated using a tetrahedron decomposition of the Brillouin zone ($6 \times 8^n$ tetrahedra in the irreducible wedge with $n = 6$ used for the density of states in order to obtain an accurate value for the Fermi energy and $n = 5$ for the susceptibility) \cite{Rath1975}. The susceptibility was calculated using the three bands crossing the Fermi energy, which are shown in Fig.~\ref{fig:suscept}(b,c).

Figures~\ref{fig:suscept}(d,e) show a decomposition of the susceptibility into total, intra-band ($n = n'$) and inter-band ($n \neq n'$) contributions in the $q_{z}=0$ and $q_{z}=\pi/c$ planes, respectively. Interestingly, we find there are {\it no} obvious trends for nesting. This is contrast to previous calculations using a tight binding model where they find a van Hove singularity very near the Fermi level~\cite{zhang2023structural}.

\section*{\label{sec:adddft}Supplementary Note 5: Additional DFT data}

\subsection{\label{sec:dftu}Energetics and magnetic moments}
As a best practice, DFT+$U$ requires careful analysis of the selected double-counting correction. Here, we track the influence of the choice of double-counting correction (AMF or FLL) on the energetics and the magnetic moments of \LNO{7} at 0 GPa and 29.5 GPa.

\begin{figure*}
    \centering
    \includegraphics[width=0.9\columnwidth]{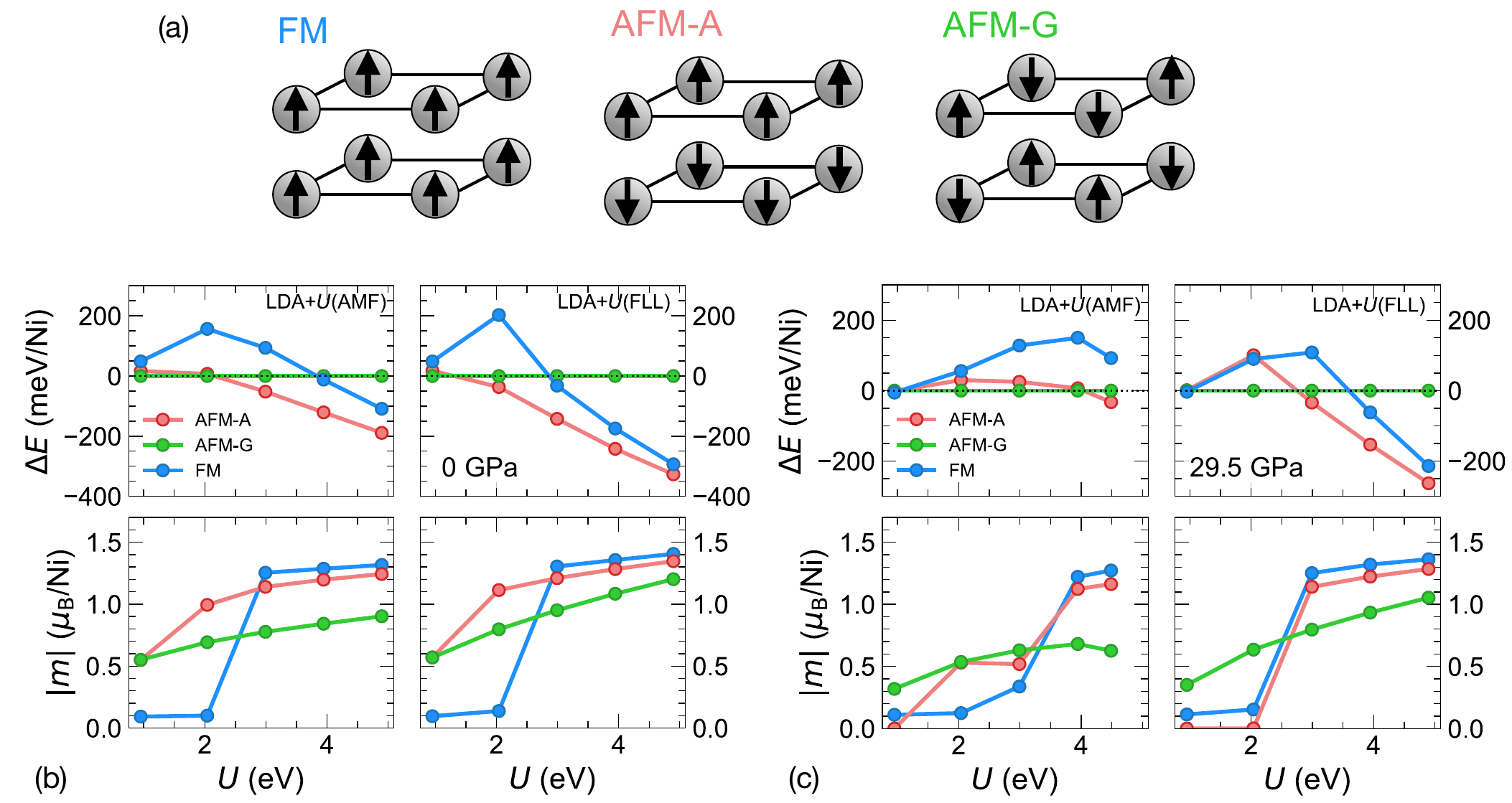}
    \caption{(a) Schematic diagram of the different magnetic orderings considered. Energetics and magnetic moments obtained within LDA+$U$ (AMF and FLL) for (b) P = 0 GPa and (c) P = 29.5 GPa. }
    \label{fig:energiesmoments}
\end{figure*}

Figure \ref{fig:energiesmoments} summarizes the systematic change in the energetics and magnetic moments within LDA+$U$ (AMF or FLL) for \LNO{7} at 0 GPa and 29.5 GPa for a range of $U$ values. Note that for all calculations the Hund's coupling $J_{\mathrm{H}}$ has been fixed to 0.7 eV. The LDA+$U$(AMF) results have been described in the main text. The data shown in Fig.~\ref{fig:energiesmoments} includes the energetics of the ferromagnetic solution as at higher values of $U$, a crossover between the AFM-A and a FM state will occur. Compared to the AMF results, we can see that the choice of FLL gives qualitatively similar energetics and moments. Importantly, a spin-state transition still occurs within LDA+$U$(FLL), but, as FLL strongly favors high-spin solutions, the range of $U$s for which the AFM-G LS state is stable is smaller relative to the LDA+$U$(AMF) scheme.

\subsection{\label{sec:highUsolutions}High-spin AFM-A solutions}
\begin{figure*}
    \centering
    \includegraphics[width=0.8\columnwidth]{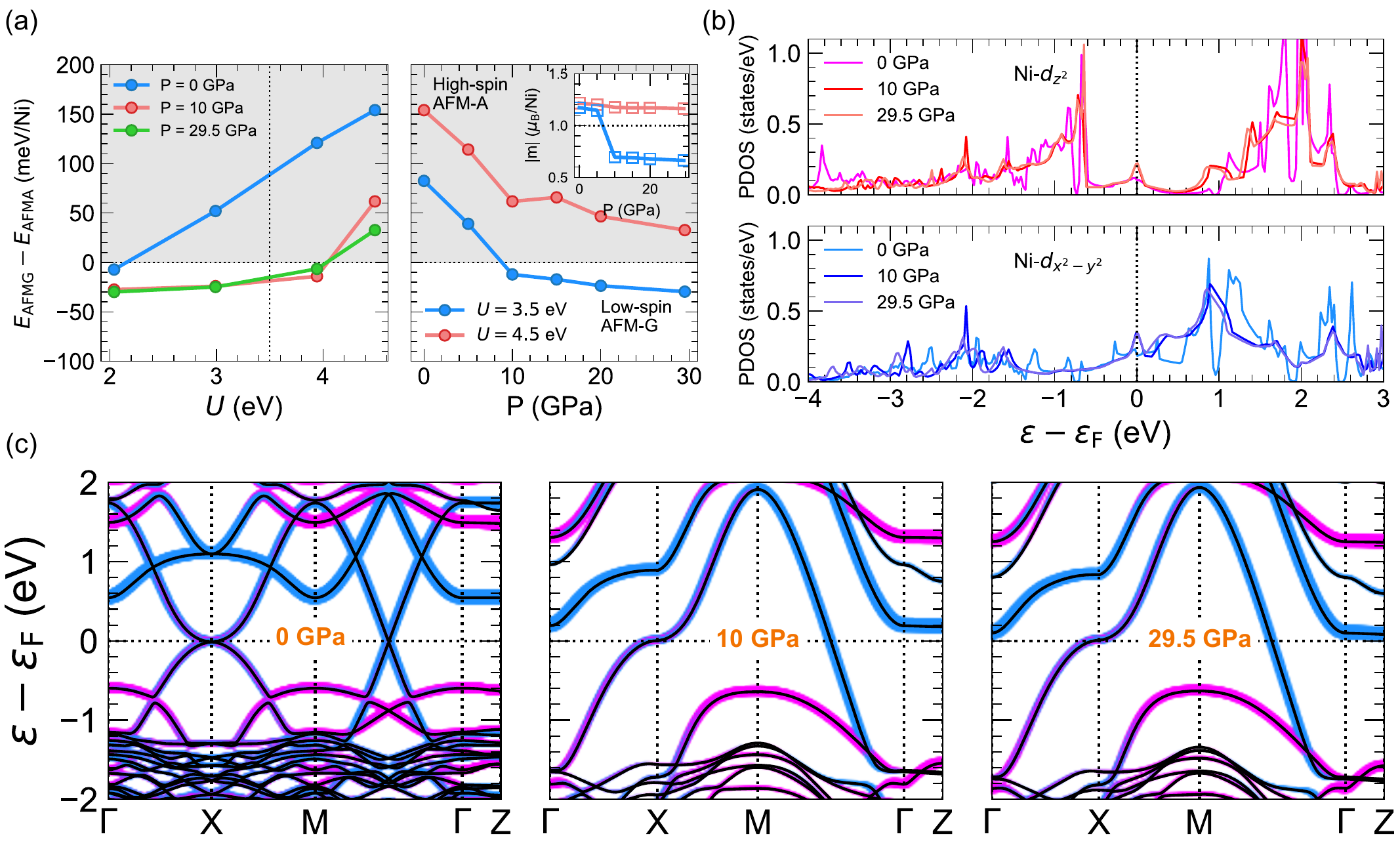}
    \caption{AFM-A high-spin solutions at P = 0 GPa ({\it Amam}), 10 GPa ({\it Fmmm}), and 29.5 GPa ({\it Fmmm}) (left to right) with LDA+$U$(AMF) ($U = 4.5$ eV, $J_{\mathrm{H}} = 0.7$ eV). (a) Same energy differences as those in Fig. 3 of the main text with $U = 4.5$ eV data showing the lack of a spin-state transition at higher values of $U$. (b) Ni-$e_{g}$ PDOS at different pressures. (c) Band structures along high-symmetry lines with Ni-$e_{g}$ orbital character highlighted. Ni-$d_{x^{2}-y^{2}}$ (Ni-$d_{z^{2}}$) is shown in blue (pink).}
    \label{fig:afma-hs}
\end{figure*}

As described in the main text, for larger values of $U$ we find that the ground state of \LNO{7} does not change with pressure and remains in the AFM-A magnetically ordered phase. The electronic structure for the AFM-A (HS) solutions at $U = 4.5$ eV ($J_{\text{H}}=0.7$ eV) for three representative pressures is summarized in Fig.~\ref{fig:afma-hs}(a,b,c) within LDA+$U$ (AMF). A common feature in all cases is the presence of a (quasi) van Hove singularity pinned near the Fermi level. This feature is also present within the FLL double-counting scheme (not shown). This suggests that these HS AFM-A solutions are unstable. Possible mechanisms to relieve this instability would require further investigation. 

\subsection{\label{sec:occ}DFT occupations and general trends}

\begin{figure*}
    \centering
    \includegraphics[width=0.7\columnwidth]{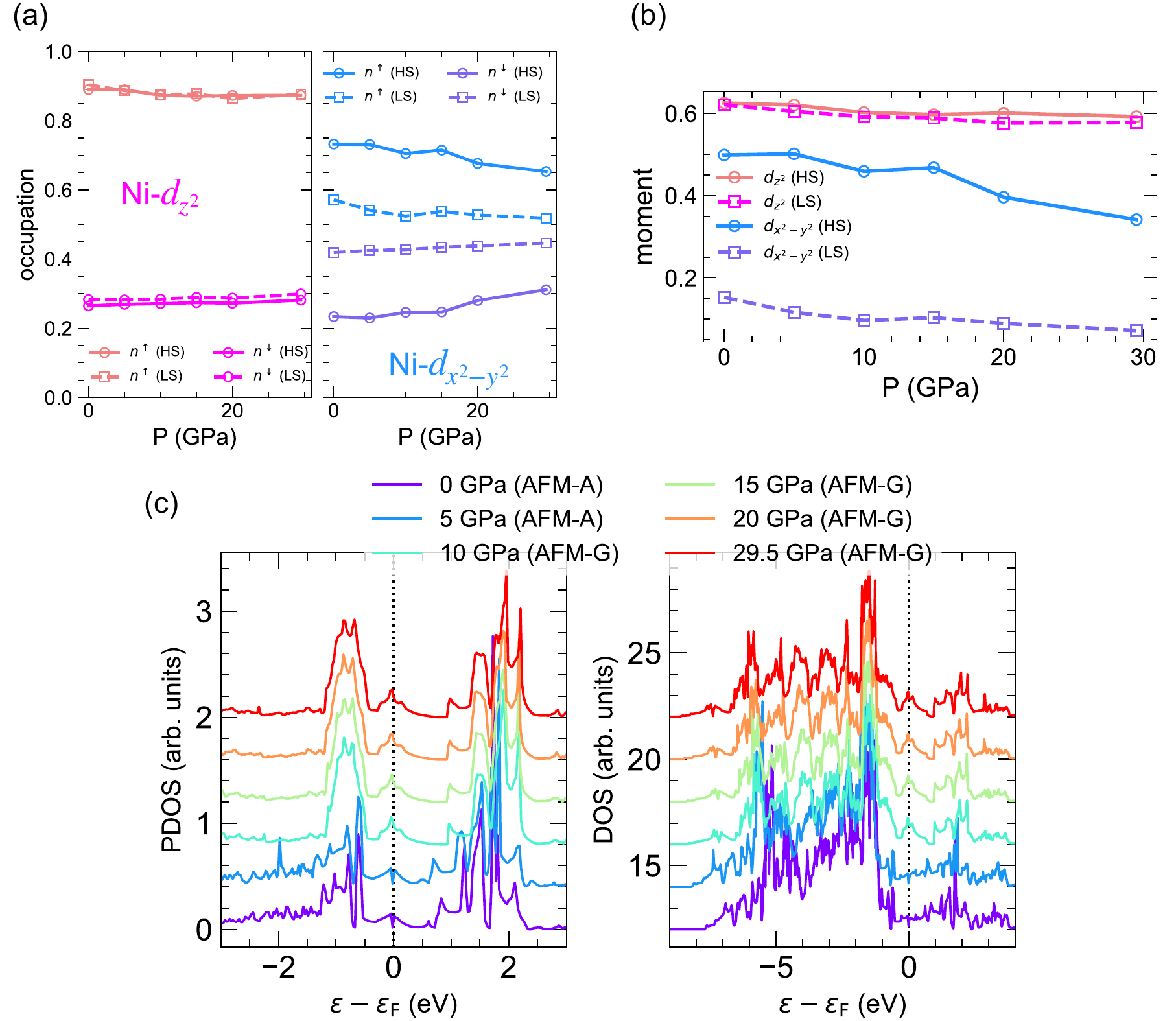}
    \caption{(a) Spin-resolved and orbital-resolved occupations ($n^{\sigma}$) and (b) moments ($n^{\uparrow}-n^{\downarrow}$) for the HS (AFM-A) and LS (AFM-G) solutions as a function of pressure. (c) Systematic trends from the ground-state electronic structure of \LNO{7} at various pressures. PDOS of the Ni-$d_{z^{2}}$ orbitals (left) and Ni($3d$)+O($2p$) DOS (right) shows that the Ni-$d_{z^{2}}$ bonding-antibonding splitting and $p$-$d$ splitting (proxy for charge-transfer energy) both increase with pressure.}
    \label{fig:adddft}
\end{figure*}

Figure~\ref{fig:adddft}(a,b) shows the spin-resolved and orbital-resolved occupations ($n^{\sigma}$) and moments ($n^{\uparrow}-n^{\downarrow}$) for the HS (AFM-A) and LS (AFM-G) solutions as a function of pressure, respectively. For the Ni-$d_{z^{2}}$ orbitals, we see that the occupations and moments change the same at all pressures when comparing the HS and LS solutions. In contrast, the occupation of the Ni-$d_{x^{2}-y^{2}}$ orbitals is very different when comparing the HS and LS solutions.

Figure~\ref{fig:adddft}(c) reveals generic trends in ground state electronic structure of \LNO{7} as a function of pressure. The Ni-$d_{z^{2}}$ bonding-antibonding splitting increases systematically with pressure as expected. From the Ni($3d$)+O($2p$) DOS, we see that the centroid of the O($2p$) bands is pushed to lower energy as pressure increases. This decreases the $p$-$d$ hybridization, thus increasing the charge-transfer energy, similar to the trends described in the main text for the non-magnetic calculations.

\clearpage
\newpage
\bibliography{ref.bib}